\documentclass[aps,preprint,amsmath,amssymb]{revtex4-1}
\usepackage{graphicx}
\newcommand{\nn}{\nonumber}
\newcommand{\bd}{\begin{document}}
\newcommand{\ed}{\end{document}}
\newcommand{\bc}{\begin{center}}
\newcommand{\ec}{\end{center}}
\newcommand{\be}{\begin{eqnarray}}
\newcommand{\ee}{\end{eqnarray}}
\newcommand{\ba}{\begin{array}}
\newcommand{\ea}{\ed{array}}
\newcommand{\strich}[1]{#1  \! \! \slash}
\newcommand{\eqn}{\global\def\theequation}
\newcommand{\sw}{sin^2 \theta_W}
\newcommand{\fbd}{f_B}
\renewcommand{\thefootnote}{\alph{footnote}}
\newcommand{\se}{\section}
\newcommand{\sse}{\subsection}
\newcommand{\bi}{\bibitem}
\def\figcap{\section*{Figure Captions\markboth
     {FIGURECAPTIONS}{FIGURECAPTIONS}}\list
     {Figure \arabic{enumi}:\hfill}{\settowidth\labelwidth{Figure 999:}
     \leftmargin\labelwidth
     \advance\leftmargin\labelsep\usecounter{enumi}}}
\let\endfigcap\endlist \relax
\def\reflist{\section*{References\markboth
     {REFLIST}{REFLIST}}\list
     {[\arabic{enumi}]\hfill}{\settowidth\labelwidth{[999]}
     \leftmargin\labelwidth
     \advance\leftmargin\labelsep\usecounter{enumi}}}
\let\endreflist\endlist \relax

\def\Journal#1#2#3#4{{#1} {{\bf #2},} {#4} {(#3)}}
\def\NCA{Nuovo Cimento}
\def\NIM{Nucl. Instrum. Methods}
\def\NIMA{{Nucl. Instrum. Methods} A}
\def\NP{{Nucl. Phys.} }
\def\NPB{{Nucl. Phys.} B }
\def\NPA{{Nucl. Phys. A}}
\def\PLB{{Phys. Lett.}  B}
\def\PL{{Phys. Lett.}}
\def\PPSA{{Proc. Phys. Soc.} A}
\def\PRP{{ Phys. Rep.}}
\def\PRL{ Phys. Rev. Lett.}
\def\PR{{Phys. Rev.}}
\def\PRD{{Phys. Rev.} D}
\def\PRC{{Phys. Rev.} C}
\def\ZP{{Z. Phys.}}
\def\ZPC{{Z. Phys. C}}
\def\EPJ{{Eur. Phys. J.}}
\def\EPJC{{Eur. Phys. J.} C}
\def\ZPA{{Z. Phys.} A}
\def\MPL{{Mod. Phys. Lett.}}
\def\MPLA{{Mod. Phys. Lett.} A}
\def\CPC{Comput. Phys. Commun.}
\def\JHEP{{J. High Energy Phys.}}
\def\JPG{{J. Phys. G.}}
\def\SJNP{Sov. J. Nucl. Phys.}
\def\NCA{ Nuovo Cimento}
\def\NIM{ Nucl. Instrum. Methods}
\def\NIMA{{ Nucl. Instrum. Methods} A}
\def\NP{{ Nucl. Phys.}}
\def\ANP{{Adv. Nucl. Phys.}}
\def\CPC{{Comput. Phys. Commun.}}

\begin{document}
\title
{\Large {\bf Study of pesudoscalar transition form factors
 within light front quark model}
}

\author{Chao-Qiang Geng$^{1,2,3}$\footnote{E-mail address: geng@phys.nthu.edu.tw}
and 
Chong-Chung Lih$^{4,3}$\footnote{E-mail address: cclih@phys.nthu.edu.tw} 
}
\affiliation{
$^1$College of Mathematics \& Physics, Chongqing University of Posts \& Telecommunications, Chongqing, 400065, China\\
$^2$Department of Physics, National Tsing Hua University, Hsinchu, Taiwan 300 \\
$^3$Physics Division, National Center for Theoretical Sciences, Hsinchu, Taiwan 300\\
$^4$Department of Optometry, Shu-Zen College of Medicine and Management,
Kaohsiung Hsien,Taiwan 452  
}

\date{\today}

\begin{abstract}
We study the  transition form factors of the pesudoscalar mesons ($\pi,\eta$ and $\eta^{\prime}$)
as functions of the momentum transfer $Q^2$ within the light-front quark model. 
We compare our results with the recent experimental data by CELLO, CLEO, BaBar and  Belle.
By considering the possible uncertainties from the quark masses, we illustrate that our predicted form 
factors can fit with all the  data, including those at the large $Q^2$ regions.

\end{abstract}


\maketitle %


Recently, the Belle collaboration~\cite{Belle} has just published its data on 
the transition form factor ($F_{\pi\gamma}$) of $\pi^{0} \to \gamma^* \gamma$, previously measured by
BaBar~\cite{BaBarPi}, CLEO~\cite{CLEO} and CELLO~\cite{CELLO} collaborations, respectively.
However, for the momentum transfer $Q^{2}$ above 10 GeV$^2$
the new data by Belle seem to be much lower than those by BaBar. 
As a result, the argument for the violation of the QCD asymptotic limit~\cite{pqcd}
is weakened despite the extensive theoretical studies in the 
literature~\cite{HOc2,BDA1,BDA2,BDA3,BDA4,BDA5,p0,p1,p2,p3,p4,p4a,p5,p6,p7,p8,p9,p10,p11,p12,p13,chpta,ldqcd,holQCD,p14}.

In addition, following the pion data~\cite{BaBarPi}, 
 the  transition form factors ($F_{(\eta,\eta')\gamma}(Q^2)$) of $\eta,\eta'\to \gamma^* \gamma$ 
have been reported by the BaBar collaboration~\cite{BaBar1} for 
$Q^{2}$ up to about $35$ GeV$^2$. 
Many theoretical works on the $\eta^{(\prime)}$ 
transition form factors have been also done~\cite{lcwf,eta1,eta2,eta3,mpa,lcqcd,LFDA,ldqcdsr1,anomalysr,eta4,Etan2}
and the results are in agreement with the data by BaBar~\cite{BaBar1}.
In particular, some of the studies  have also tried to combined the analyses on the three pesudoscalar mesons of $\pi^0$, $\eta$ and
$\eta^{(\prime)}$ to fit all data simultaneously.

Motived by the Belle data, in this note we would like to re-examine the  transition form factor of the pion along with
those of $\eta$ and $\eta^{\prime}$ within the light front quark model (LFQM)
by including uncertainties of  quark masses to check if we can accommodate all the data. Similar studies in other QCD models
have  been performed recently  in Refs.~\cite{n1,n2,n3,n4,n5}.

We will use the phenomenological light front (LF) meson wave function~\cite{lfqm1,lfpi} 
to evaluate $Q^2 |F_{\eta^{(\prime)}}(Q^2)|$ in all allowed kinematic region. 
The LF wave function 
can be constructed by the simple structure of the meson constituent in terms of   a quark-antiquark ($Q\bar{Q}$) pair~\cite{lfpi}. 
The decay 
amplitude of $Q\bar{Q} \to \gamma^*\gamma^*$ with Lorentz structure is given by~\cite{vex3}:
\be A(Q\bar{Q}(P)\to
\gamma^*(q_1,\epsilon_1)~\gamma^*(q_2,\epsilon_2))
=ie^{2}F_{Q\bar{Q}}(q^2_1,q^2_2)~\varepsilon_{\mu\nu\rho\sigma}~\epsilon^\mu_1
 ~\epsilon^\nu_2 ~q^\rho_1 ~q^\sigma_2\,, \label{def}
\ee
where $F_{Q\bar{Q}}(q^2_1,q^2_2)$ 
is a symmetric function under the interchange of $q^2_1$ and $q^2_2$,
which
 can be found to be~\cite{lfqm1,lfpi}
  \be
F_{Q\bar{Q}}(q_{1}^{2},q_{2}^{2}) &=&-4 \sqrt{N_{c}\over 3}
        \int \frac{dx\,d^{2}k_{\bot }}{2\left( 2\pi \right) ^{3}}\Phi_{Q\bar{Q}}\left( x,k_{\bot }^{2}\right) 
        {e^{2}_{Q}\over 1-x}\frac{m_{Q}+(1-x)m_{Q} k_{\bot}^{2}\Theta}{x(1-x)q_{2}^{2}-m_{Q}^{2}-k_{\bot }^{2}}
       +(q_2 \leftrightarrow q_1)  \,,
\label{fffv}
\ee
with
\be
\Theta = {\frac{1}{\Phi_{Q\bar{Q}}(x,k_{\bot}^2) }} {\frac{d\Phi_{Q\bar{Q}}(x,k_{\bot}^{2})}{%
dk_{\bot}^2}} \, ,
\ee
where $N_c=3$ is the number of colors, $e_{Q}=2/3 (-1/3)$ for $Q=u(d,s)$, 
$m_Q$ is the quark mass and $\Phi_{Q\bar{Q}} (x,k_{\bot}^2)$ is
the meson wave function, defined by
\be
\Phi_{Q\bar{Q}} (x,k_{\bot}^2) &=& \sqrt{ N_c{\frac{x(1-x) }{2 M_0^2}}} \phi_{Q\bar{Q}}(x, k_\perp)\,,
\ee
where
\be
\label{res}
\phi_{Q\bar{Q}}(x, k_\perp) &=& N \sqrt{{\frac{dk_{z}}{dx}}}\exp \left( -{\frac{\vec{k}^{2}}{%
2\omega_{Q\bar{Q}}^{2}}}\right)\,,\\
M_0^2&=&{ m_{Q}^2+k_\bot^2\over x}+{ m_{Q}^2+k_\bot^2\over
1-x}\, ,
\label{n6}
\ee
with $N = 4 ( \pi/\omega_{Q\bar{Q}}^{2})^\frac{3}{4}$, $\vec k =
(k_{\bot}, k_z)$,  $k_z=(x+1/2)M_0$,
and $\omega_{Q\bar{Q}}$ the parameter related to the physical size of the pseudoscalar meson $(P=\sum Q\bar{Q})$
in the wave function.
If $q_1$ or $q_2$ is on mass shell,
the form factor of $Q\bar{Q} \to \gamma^{*} \gamma$ can be written as
\be
F_{Q\bar{Q} \to \gamma^* \gamma}(Q^2,0)&=&-4\sqrt{N_{c}\over 3}
       \int \frac{dx\,d^{2}k_{\bot }}{2\left( 2\pi \right) ^{3}} {e^{2}_{Q} \Phi_{Q\bar{Q}}
        \left( x,k_{\bot }^{2}\right)\over 1-x}
\bigg\{ \frac{m_{Q}+(1-x)m_{Q} k_{\bot}^{2}\Theta}{x(1-x)Q^{2}-m_{Q}^{2}-k_{\bot }^{2}}
        \nn \\
 &&~~~-\frac{m_{Q}+(1-x)m_{Q} k_{\bot}^{2}\Theta}{m_{Q}^{2}+k_{\bot }^{2}} \bigg\} ,~\
\label{realff}
\ee
where $Q^2=q_1^2$ or $q_2^2$ is the momentum transfer. From Eq.~(\ref{realff}), by summing up the relevant Fock states
we obtain the transition for the transition form factors of  $P \to \gamma^* \gamma$ to be 
\be
F_{P \to \gamma^* \gamma}(Q^2)\equiv F_{P \gamma}(Q^2) &=&\sum F_{Q\bar{Q} \to \gamma^* \gamma}(Q^2,0)\,.
\label{realff1}
\ee


For the $\pi^0$ meson, we use $|\pi^0\rangle=|u\bar{u}-d\bar{d}\rangle/\sqrt{2}$ and  
$m_u=m_d=m_q$.
The states of $\eta$ and $\eta'$ can be expressed in terms of 
  the two orthogonal states of $|\eta_{q}\rangle$ and $|\eta_{s}\rangle$,
parameterized as~\cite{phi0,phi1,phi2,phi3}
\be
\left(
\begin{array}{c}
|\eta\rangle  \\
|\eta'\rangle
\end{array}
\right)\,=
\left(
\begin{array}{cc}
\cos\phi & -\sin\phi  \\
\sin\phi & \cos\phi
\end{array}
\right)\,
\left(
\begin{array}{c}
|\eta_{q}\rangle  \\
|\eta_{s}\rangle
\end{array}
\right)\,,
\label{mix}
\ee
where $|\eta_{q}\rangle=|u\bar{u}+d\bar{d}\rangle/\sqrt{2}$
and $|\eta_{s}\rangle=|s\bar{s}\rangle$. The mixing angle
has been studied in various decay processes and constrained to be 
$\phi\simeq 37^{\circ} \sim 42^{\circ}$~\cite{phi3}.
Under this scheme, the valence states of $\eta^{(\prime)}$
can be written as:
\be
|\eta\rangle&=&\cos\phi\frac{|u\bar{u}+d\bar{d}\rangle}{\sqrt{2}}-
\sin\phi|s\bar{s}\rangle\,,\nonumber \\
|\eta'\rangle&=&\sin\phi\frac{|u\bar{u}+d\bar{d}\rangle}{\sqrt{2}}+
\cos\phi|s\bar{s}\rangle\,.
\ee
Consequently, 
the  transition from factors of $\eta^{(\prime)}\to \gamma^*\gamma$ have the forms
\be
F_{\eta\gamma}&=&\cos\phi F_{\eta_{q}}-\sin\phi F_{\eta_{s}}\,, 
\nonumber\\
F_{\eta'\gamma}&=&\sin\phi F_{\eta_{q}}+\cos\phi F_{\eta_{s}}\,.
\label{Angle}
\ee

To numerically calculate the transition form factors of 
$P\to \gamma^*\gamma$, we need to
specify the parameters appearing in $\phi_{Q\bar{Q}}(x,k_\bot)$. To
constrain the quark masses of $m_{u,d,s}$ and the meson scale
parameters of $\omega_{Q\bar{Q}}$ in Eq. (\ref{res}), 
we  use the branching ratios of $P \to 2\gamma$ and the decay constants of the $Q\bar{Q}$ states,
defined by
\be
{\cal B}(P \to 2\gamma)&=& \frac{(4\pi \alpha)^{2}}{64\pi \Gamma_P} m_P^{3} |F(0,0)_{P \to 2\gamma}|^2 \,,
\ee
and 
\be
f_{Q\bar{Q}}&=&\,4{\sqrt{N_c}\over\sqrt{2}}\int {dx\,d^2k_\perp\over 2(2\pi)^3}\,\phi_{Q\bar{Q}}(x,
k_\perp)\,{m_Q\over\sqrt{m_Q^{2}+k_\perp^2}}\,,
\label{fq}
\ee
respectively,
where $Q=q$ or $s$ denotes the quark in the Fock state. 
Explicitly, we use~\cite{pdg}
\be
{\cal B}(P \to 2\gamma)=\,(98.832\pm0.034)\,,~~(39.30\pm0.20)\,,~~(2.12\pm0.14)\%\,\,, \label{br2r}
\ee
which lead to  $|F(0,0)_{P \to 2\gamma}|\equiv|F_{P\gamma}(0)|=0.274$, $0.260$ and  $0.341$ in $GeV^{-1}$
for $P=\pi^0$, $\eta$ and $\eta^{\prime}$, respectively. 
For the decay constants, we take~\cite{lcqcd}
\be
f_{\pi}&=&\,132\,,~~f_{q\bar{q}}=\,140\,,~~f_{s\bar{s}}=\,168\,{\rm MeV}\,.
\label{fpl}
\ee

\begin{figure}[htbp]
\includegraphics*[width=4in,height=3in,angle=0]{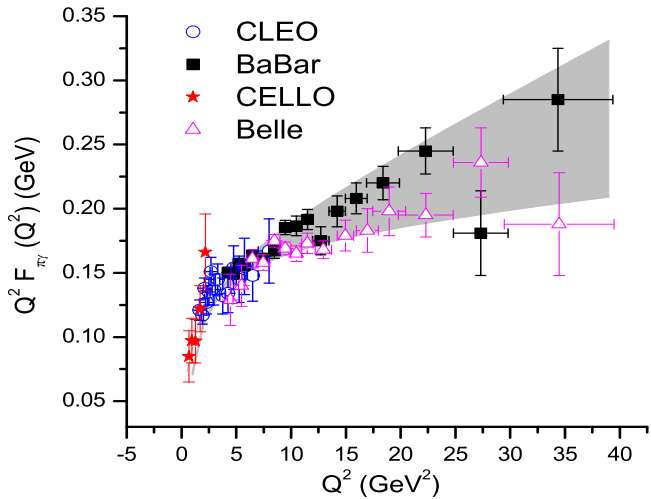}
\caption{ (Color online) $Q^2 F_{\pi\gamma}(Q^2)$ as a function of $Q^2$ in the LFQM 
with  $m_q=0.22\sim0.30$ GeV.}
\label{Fig1}
\end{figure}

 To illustrate the pion transition form factor, we have to specify the up and down quark masses.
 In our previous study in Ref.~\cite{lfpi}, we have fixed $m_q=0.24$ GeV ($q=u,d$). 
 To fit all the experimental data including the new data from Belle, 
 we could like to include the uncertainty from the quark masses.  
Explicitly, we revise our input with a possible range of $m_q$, $i.e.$
$m_q=0.22\sim 0.30$  GeV. 
As a result, we can derive  various meson scale parameters of $\omega_{\pi}$ from 
the pion decay constant and $F_{\pi\gamma}(0)$ from
the decay rate of $\pi^0\to\gamma\gamma$.
In Fig.~\ref{Fig1}, we show the $Q^2$ dependence of the $\pi^0$ transition form 
factor $Q^2 F_{\pi\gamma}(Q^2)$ in the LFQM (gray band) with $m_q=0.22\sim0.30$ GeV, where we have also plotted the experimental
data of BaBar~\cite{BaBarPi},  Belle~\cite{Belle}, CELLO~\cite{CELLO} and CLEO~\cite{CLEO}. 
Note that the upper  and lower edges of the gray band in Fig.~\ref{Fig1} correspond to $m_q=0.30$ and $0.22$  GeV, respectively.
 From the figure, we see that  either the experimental data by CELLO, CLEO
and BaBar or those by CELLO, CLEO and Belle can be simultaneously fitted well in the LFQM. 
 We emphasize that to reproduce the BaBar (Belle) high-$Q^2$ tail, 
the higher (lower) quark mass is required.

\begin{figure}[htbp]
\includegraphics*[width=4in,height=3in,angle=0]{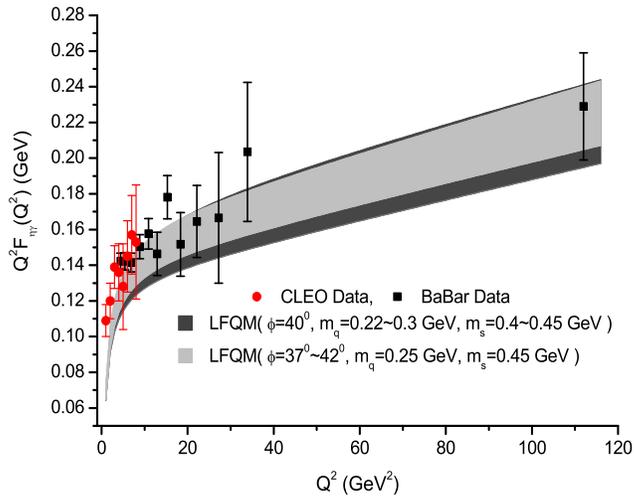}
\caption{ (Color online) $Q^2 F_{\eta\gamma}(Q^2)$ as a function of $Q^2$ in the LFQM, where the dark-gray 
band represents the inputs of $m_q=0.22\sim0.30$, $m_s=0.40\sim0.45$ in GeV and $\phi=40^\circ$,
while the light-gray one stands for those of $m_q=0.25$, $m_s=0.45$ GeV and  $\phi=37\sim42^\circ$.
}
\label{Fig2}
\end{figure}

\begin{figure}[htbp]
\includegraphics*[width=4in,height=3in,angle=0]{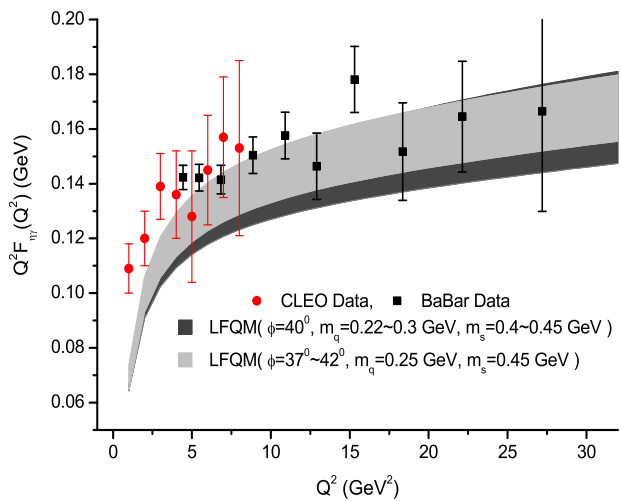}
\caption{Legend is the same as Fig.~\ref{Fig2} but for the lower $Q^2$ region of $Q^2<30$ GeV$^2$.
}
\label{Fig3}
\end{figure}

\begin{figure}[htbp]
\includegraphics*[width=4in,height=3in,angle=0]{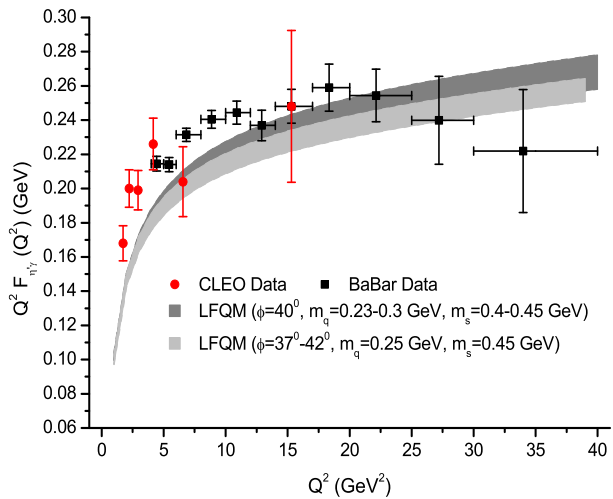}
\caption{Legend is the same as Fig.~\ref{Fig2} but for $\eta^\prime$.} 
\label{Fig4}
\end{figure}

In our numerical calculations of $\eta$ and $\eta^\prime$, 
the first term in Eq.~(\ref{realff}) dominates for the lower region of $Q^2$
and thus, it can be used to describe the experimental data of CLEO~\cite{cleo1} and 
BaBar~\cite{BaBar1} with $Q^2$ $\leq$ 10 GeV$^2$. 
The second one in Eq.~(\ref{realff}), related to the non-valence quark contributions, 
is quite small for a small $Q^2$. In general, this term can be neglected in the low $Q^2$ region,
but it may enhance the form factors of $Q^2 F_{\eta^{(\prime)}\gamma}$ at the high values of $Q^2$.
Hence, we will take into account this term in our calculations.
Similar to the pion case, we will also consider the uncertainties 
from the quark masses. Explicitly, we use 
$m_q=0.22\sim0.30$ and  $m_s=0.40\sim0.45$ in GeV. Moreover, we will  examine a possible range
of $\phi=37\sim42^\circ$ for the mixing angle in Eq.~(\ref{Angle}).
In  Fig.~\ref{Fig2},
we show our results for the $Q^2$ dependence of the $\eta$  transition form factor in terms of $Q^2 F_{\eta\gamma}(Q^2)$,
where the dark-gray band represents the inputs of $m_q=0.22\sim0.30$, $m_s=0.40\sim0.45$ in GeV and $\phi=40^\circ$,
while the light-gray one stands for those of $m_q=0.25$, $m_s=0.45$ GeV and  $\phi=37\sim42^\circ$.
The enlarged figure of Fig.~\ref{Fig2} for the lower $Q^2$ region of $Q^2<30$ GeV$^2$ is given in Fig.~\ref{Fig3}.
In Fig.~\ref{Fig4}, we draw $Q^2 F_{\eta^{\prime}\gamma}(Q^2)$ as a function of $Q^2$, where 
the dark-gray and light-gray bands represent the inputs of $m_q=0.22\sim0.30$, $m_s=0.40\sim0.45$ GeV and $\phi=40^\circ$ and
 $m_q=0.25$, $m_s=0.45$ GeV and  $\phi=37-42^\circ$, respectively.

As shown in Figs.~\ref{Fig2}-\ref{Fig4}, our results for $Q^2 F_{\eta^{(\prime)}\gamma}(Q^2)$ are in good agreement with the experimental data.
Note that the upper (lower) edges of the dark-gray bands in Figs.~\ref{Fig2}-\ref{Fig4} correspond to $m_q=0.30\,(0.25)$ and $m_s=0.45\,(0.4)$ GeV,
while those of the yellow bands $\phi=37^\circ\,(42^\circ)$. 
We remark that the form factors $Q^2 F_{\eta^{(\prime)}\gamma}$ increase (decrease) with quark masses $m_q$ 
(the mixing angle $\phi$), whereas the effect from the uncertainty from $m_s$ is small due to the small quark charge.
It is interesting to point out that the form factors can be better fitted for
a larger $m_q$ with a fixed  $\phi$ or $\phi=40^\circ$ with a fixed $m_q$  in the lower $Q^2$ region. 
 

\ \ \
In summary, motivated by the recent experimental measurements,
we have shown the transition form factors of 
$\pi^0$, $\eta$ and $\eta^{\prime}\to \gamma^*\gamma$ as functions of the momentum transfer $Q^2$
within the LFQM.
We have recalculated $F_{\pi\gamma}(Q^2)$ by considering
the allowed possible range of $m_q=0.22\sim 0.30$  GeV. We have illustrated that
out result of the pion transition form factor in the LFQM can fit  either the experimental data by CELLO, CLEO
and BaBar or those by CELLO, CLEO and Belle for a fixed quark mass.
 In particular, we have found that to reproduce the BaBar and the recent Belle high-$Q^2$ tails, 
the higher and lower quark masses are needed, respectively.
With the same set of model parameters as the pion, we have also studied the form factors of $F_{\eta^{(\prime)}\gamma}$ 
by considering the possible ranges of the quark masses:
$m_q=0.22\sim0.30$ and $m_s=0.40\sim0.45$ in GeV and the $\eta-\eta^\prime$ mixing angle:
$\phi=37\sim42^\circ$, and we have found that  our results agree well with the CLEO and BaBar data 
in the $\eta$ and $\eta^\prime$ cases.\\

This work was partially supported by National Center of Theoretical
Science, National Science Council (NSC-97-2112-M-471-002-MY3, NSC-98-2112-M-007-008-MY3 and
NSC-101-2112-M-007-006-MY3) 
 and SZL-10104009.\\

\end{document}